\documentclass[final]{aipproc}
\layoutstyle{6x9}

\usepackage{graphicx}% Include figure files

\newcommand{\onepion}{one pion~}

\newcommand{\GeV}{\; \mathrm{GeV}}

\begin{document}

\title{Pion production in the MiniBooNE}
\classification{13.15.+g, 25.30.Pt}
\keywords{neutrino reactions, nuclear effects, pion production}
\author{O. Lalakulich}{
  address={Institut f\"ur Theoretische Physik, Universit\"at Giessen, Germany}
}
\author{K. Gallmeister}{}
\author{T. Leitner}{}
\author{U. Mosel}{}

\begin{abstract}
We investigate one pion production processes within the Giessen Boltzmann--Uehling--Uhlenbeck (GiBUU)
coupled channel transport model.
Our calculations  for integrated and differential cross sections
for realistic experimental neutrino fluxes 
are compared to the data recently provided by the MiniBooNE collaboration.
\end{abstract}

\date{\today}

\maketitle

\section{Introduction}

Recently the MiniBooNE and K2K collaborations have published data on
charged~\cite{AguilarArevalo:2010bm} and neutral~\cite{AguilarArevalo:2010xt} pion production
in CC neutrino scattering.  For the first time the pion energy and angle observables were presented.
Generally, these experiments  report cross sections which are noticeably higher
than those expected from any conventional theoretical approach \cite{Dytman:2009zzb}.

In this paper we investigate by detailed comparisons with data on \onepion production 
how far a realistic impulse approximation model of these processes can go in explaining all the available data. 
For this purpose  we employ the GiBUU model developed
as a transport model for nucleon-, nucleus-,  pion-, and electron- and neutrino-induced collisions from
hundreds MeV up to tens GeV. 
Thus, the model allows one to study various types of processes on nuclei within a unified description.  
This is particularly important for broad-band neutrino experiments which inherently sum over many different reaction types. 
The code is written in modular FORTRAN and is available for download as open source \cite{gibuu}.
Our results for QE scattering and NC pion production are presented in \cite{Leitner:2009de}.

\section{GiBUU transport model}

The MiniBooNE energy flux \cite{AguilarArevalo:2008yp} has an average neutrino energy $0.7 \GeV$,
with the flux becoming very small above $2\GeV$.
Several different channels are important for describing the neutrino interactions at these energies:
quasi-elastic scattering, resonance production (with the $\Delta$ [$P_{33}(1232)$] giving the biggest contribution)
and non-resonant \onepion production. All of them are described within the GiBUU model, the details are given
in \cite{Buss:2011mx}.
Our elementary input is described in \cite{Leitner:2008ue}.
We emphasize, that in describing QE scattering  the world average value for the nucleon the axial mass
$M_A=0.999 \GeV$~\cite{Kuzmin:2007kr} is used; in describing resonance production and background processes
parameters were tuned to the ANL data.

Nuclear targets are modeled as follows.
The struck nucleus is considered  as a collection of off-shell nucleons.
Each nucleon is bound in a mean-field potential, which on average describes the many-body interactions with the other
nucleons. 
The phase space density of nucleons is treated within a local Thomas-Fermi approximation.
In the impulse approximation, an incoming lepton interacts with a single bound nucleon, with the
interaction vertex being the same as in the case of a free nucleon. Thus, the calculations 
do not contain any so-called 2p-2h interactions \cite{Nieves:2011pp}.

In the GiBUU code it is also possible to take into account the nuclear medium corrections to the resonance 
widths and spectral functions. In this paper, however, they are neglected. 
Final State Interactions (FSI) are implemented by solving the semi-classical Boltzmann-Uehling-Uhlenbeck (BUU) equation.
It describes the dynamical evolution of the phase space density for each particle species
under the influence of the mean field potential, introduced in the description of
the initial nucleus state. Equations for various particle species are coupled through this mean field and
also through the collision term. This term explicitly accounts for changes in
the phase space density caused by elastic and inelastic collisions between particles.

FSI decrease the cross sections as well as significantly modify
the shapes of the final particle spectra. Such modification was experimentally 
observed in  photo-pion production~\cite{Krusche:2004uw,Buss:2006yk,Mertens:2008np} and is described 
by the GiBUU with a high accuracy.

For a detailed review of the GiBUU model see \cite{Buss:2011mx}.

\section{One pion production and the origin of pions.}
\label{miniboone}

\begin{figure}[hbt]
\includegraphics[width=0.8\linewidth]{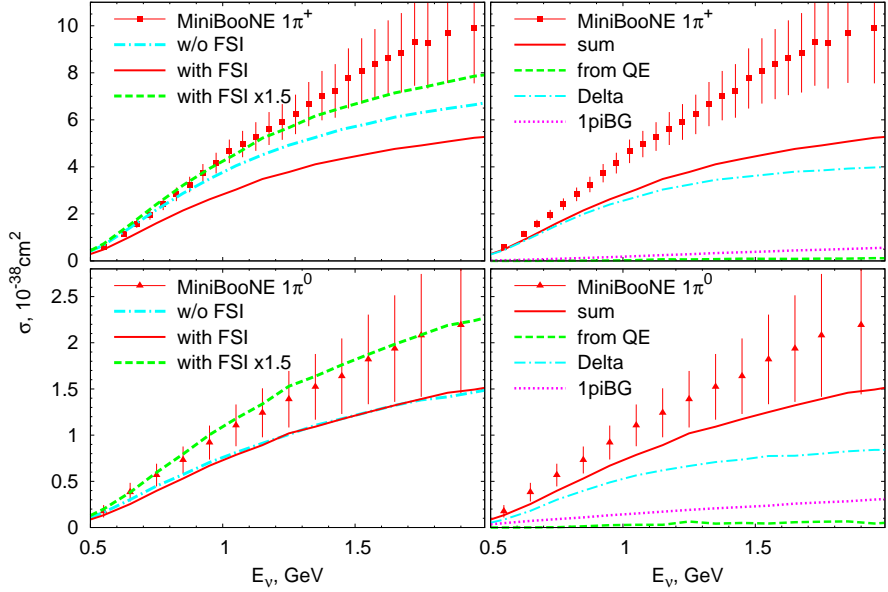}
\caption{(Color online) Integrated cross section for MiniBooNE $1\pi^+$(upper panel) and
$1\pi^0$(lower panel) CC production versus neutrino energy.
Data are from~\protect\cite{AguilarArevalo:2010bm,AguilarArevalo:2010xt}.
The panels on the right show the composition of the calculated cross sections.}
\label{fig:MB-lepton-Enu-QEDelta}
\end{figure}

Here we present our calculations on  a $CH_2$ target and compare the results with the data
from the MiniBooNE experiment \cite{AguilarArevalo:2010bm,AguilarArevalo:2010xt}.
Integrated cross sections versus neutrino energy for the charged current $1\pi^+$ and $1\pi^0$ production
are  shown in Fig.~\ref{fig:MB-lepton-Enu-QEDelta}.
As in the MiniBooNE experiment, the \onepion events are defined as ``observable \onepion production'',
i.e. events  with one pion of a given charge and no other pions in the final state,
regardless of which particles  were produced in the initial neutrino vertex.

The left upper panel in Fig.~\ref{fig:MB-lepton-Enu-QEDelta} shows the results for $\pi^+$ production.
Comparison of the curves with and without FSI shows that the FSI do not change the energy-dependence of the curves.
The curve with FSI (solid line) lies 20\% below the curve without FSI (dash-dotted line).
This reduction is mainly caused by the $\Delta$ absorption through $N \Delta \to NN$ scattering.
Charge exchange process $\pi^+ n \to \pi^0 p$ also depletes the dominant $1\pi^+$ channel.
The side feeding from the reverse
process gives minor relative contribution, because the initial $\pi^0$ production cross section is around 5 times lower.
For the 1$\pi^0$ cross section in the left lower panel the same side feeding processes
increase the cross section. This is, however, compensated by absorption, charge exchange to the $\pi^-$ channel through
$\pi^0 n \to \pi^- p$ and other channels such as $\pi N \to \Sigma K, \Lambda K$. The overall effect is nearly the same
cross section with and without FSI.

The right upper panel shows the origin of the $1\pi^+$ events. The most of them (dash-dotted line)
come form the initial  Delta resonance production and its following decay. Some  events (dotted line) are
background ones. A minor amount comes from the initial QE vertex (long-dashed line),
which is only possible due to FSI, when the outgoing proton is rescattered.
Here the main contribution is from  the $p N \to N' \Delta \to N' N^{''} \pi$ reaction.
The right lower panel shows the origin of the $1\pi^0$ events. Here the background processes and the FSI
play an even bigger role.

We now turn to a discussion of a comparison with experiment. Already the curves \emph{without}
any FSI  lie considerably ($\approx 25$\%) \emph{below} the data,
those with FSI included (solid line) are nearly by a factor of about 1.5 (at $1 \GeV$)
below the experimental data;
at higher neutrino energies these discrepancies become even larger since the experimental cross
sections rise more steeply with energy than the  calculated pion cross sections.
At 2 GeV the discrepancy between the results with FSI and the data amounts to a factor of $\approx 2$!
For 1$\pi^0$ production  one observes the similar result: the data are considerably higher than our calculated values. 
The slope of the curve, however,
is in agreement with the experiment, as one can conclude from comparison
of the ``with FSI'' curve multiplied by a factor of 1.5 (long-dashed curve) with the data.

\begin{figure}[htb]
\begin{minipage}[t]{0.48\linewidth}
\includegraphics[width=\linewidth]{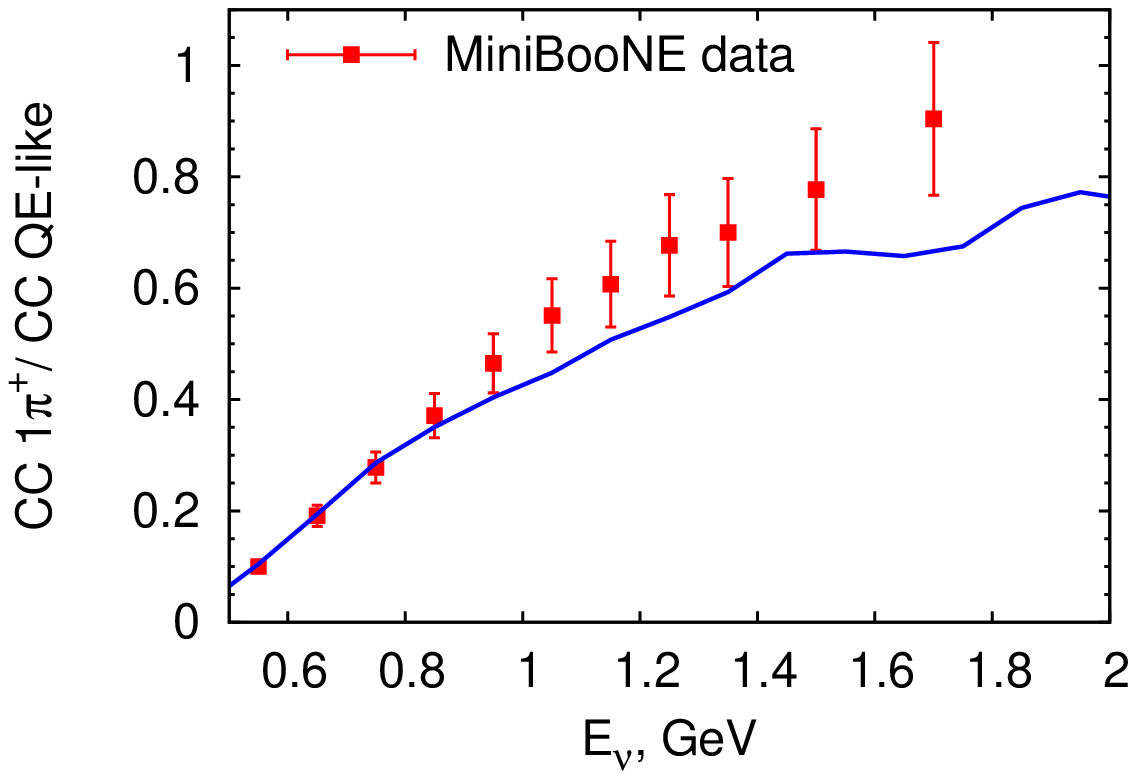}
\caption{(Color online) (Left panel) Observed single 1$\pi^+$/QE-like cross sections ratio
for CC neutrino scattering versus neutrino energy.
Data are from~\protect\cite{AguilarArevalo:2009eb}.
(Right panel) Integrated cross section for MiniBooNE $\pi^-$ CC production versus neutrino energy.}
\label{fig:MB-CC1pi-to-QElike}
\end{minipage}
\hfill
\begin{minipage}[t]{0.48\linewidth}
\includegraphics[width=\linewidth]{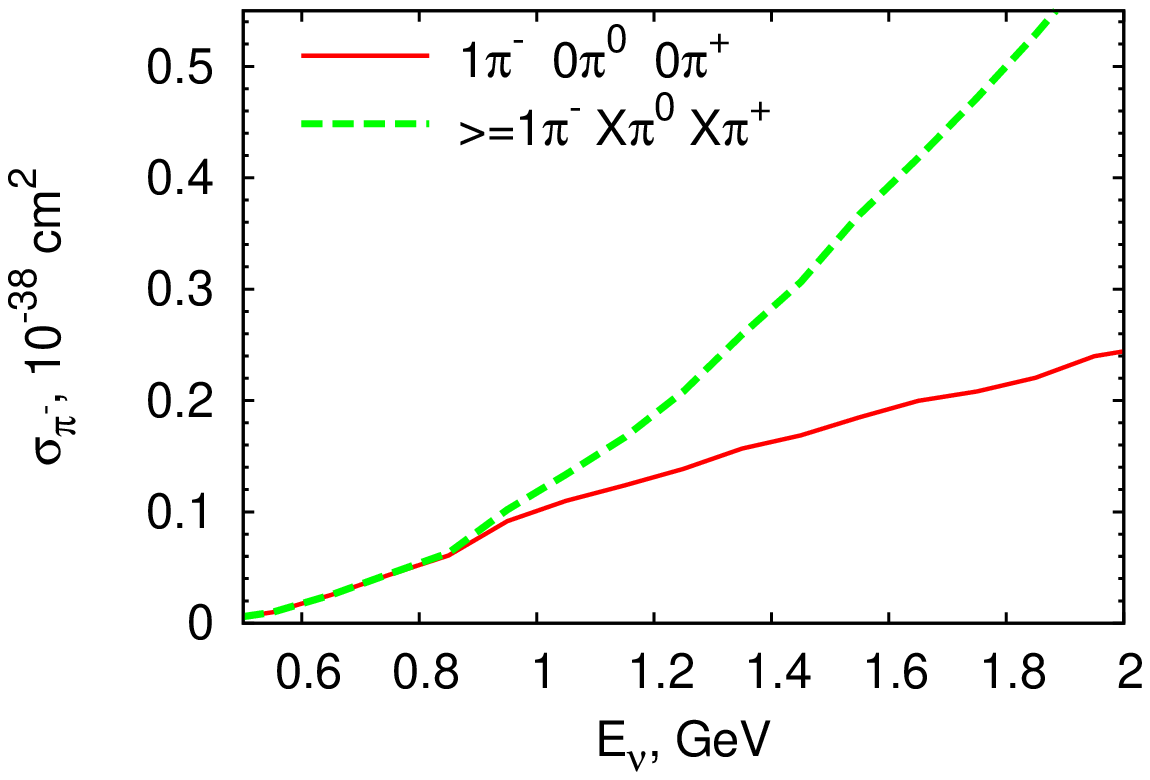}
\end{minipage}
\end{figure}

An observable that is approximately free of the uncertainties of the neutrino flux is the ratio
of pion production to QE-like scattering at fixed neutrino energy. Here, of course, an
additional uncertainty enters because the energies for QE and for
\onepion production are being reconstructed independently and from
different underlying assumptions. The errors in the two independent energy-determinations thus add.

This is essential since our calculations for the ratio of $1\pi^+$ to
QE-like cross sections shown in the left panel of Fig.~\ref{fig:MB-CC1pi-to-QElike}
gives a reasonably good agreement with the data up to about 1 GeV neutrino
energy and even above that the discrepancy amounts to only
about 10\%.
This uncertainty is well within the limits of
the combined errors of flux and energy reconstruction. Note here that the experimental QE-like cross section most probably contains contributions from 2p-2h excitations while the theoretical does not. Including those in the calculations would lower the ratio by about 25\% at 1 GeV.

Our prediction for the $\pi^-$ cross  section are shown in the right panel of Fig.~\ref{fig:MB-CC1pi-to-QElike}.
Without FSI the production of $\pi^-$ is negligible, a few events may only come from higher
mass resonances in the processes like $R^+ \to p \rho^0 \to p \pi^+ \pi^- $ or
$R^+ \to p \Delta^0 \to p p \pi^-$ (the curves are not shown because they would be indistinguishable from zero).
Cross sections for both $1\pi^-$ (solid curve), which is defined as one $\pi^-$ and no other pions, and multi-$\pi^-$,
which is defined as at least one  $\pi^-$ and any number of pions of other charges, are shown.
The cross sections for $\pi^-$ production appear to be around 20  times smaller than those
for $1\pi^+$. However, with the current statistics of the MiniBooNE experiment they should
be measurable.

\section{Lepton spectra}

\begin{figure}[!hbt]
\begin{minipage}[c]{0.48\textwidth}
\includegraphics[width=\textwidth]{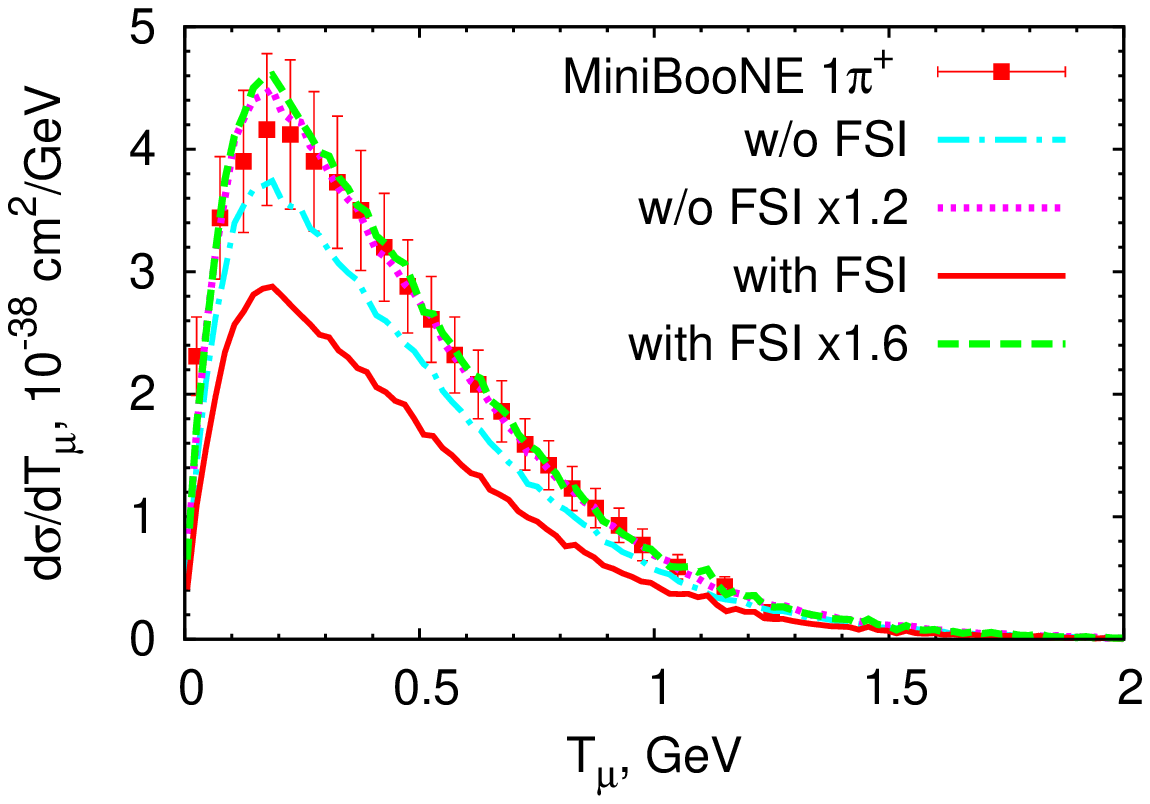}
\end{minipage}
\hfill
\begin{minipage}[c]{0.48\textwidth}
\includegraphics[width=\textwidth]{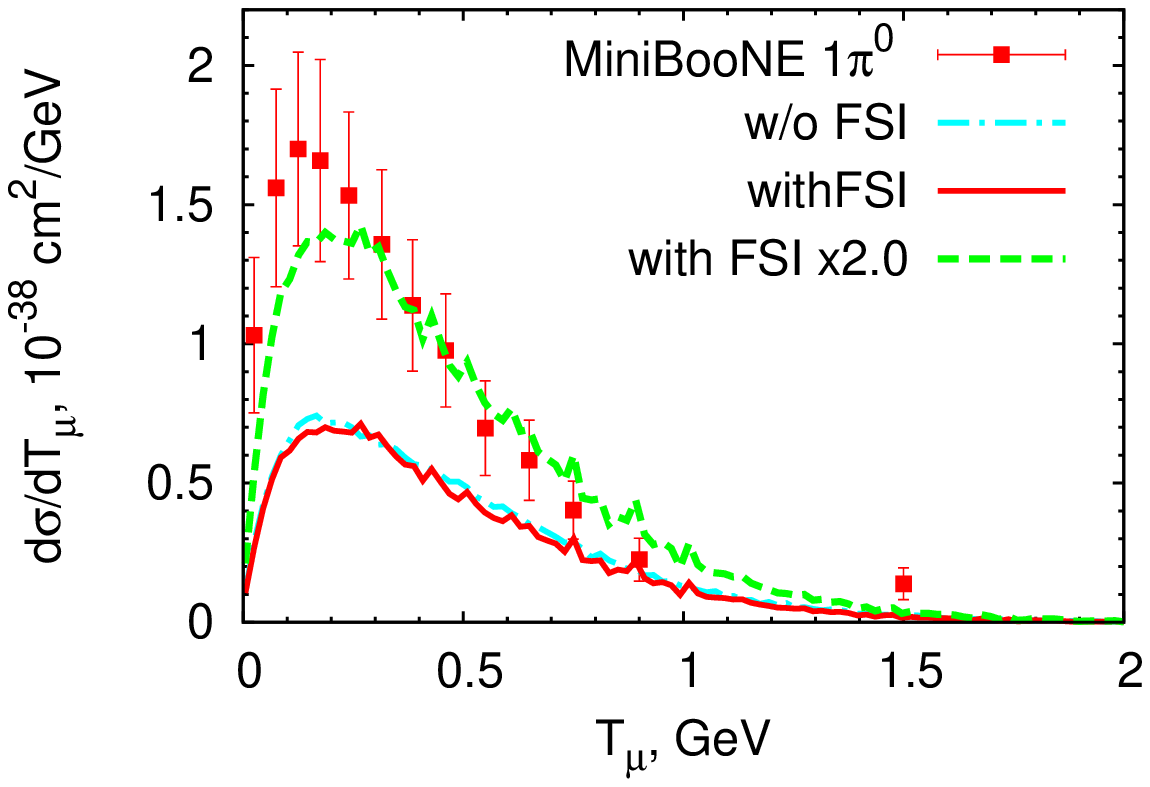}
\end{minipage}
\caption{(Color online) Distribution of the outgoing muons in their kinetic
energy for MiniBooNE CC 1$\pi^+$(left panel) and 1$\pi^0$ (right panel) neutrino production.
Data are from~\protect\cite{AguilarArevalo:2010bm,AguilarArevalo:2010xt}.}
\label{fig:MB-lepton-Ekin}
\end{figure}

\begin{figure}[!hbt]
\begin{minipage}[c]{0.48\textwidth}
\includegraphics[width=\textwidth]{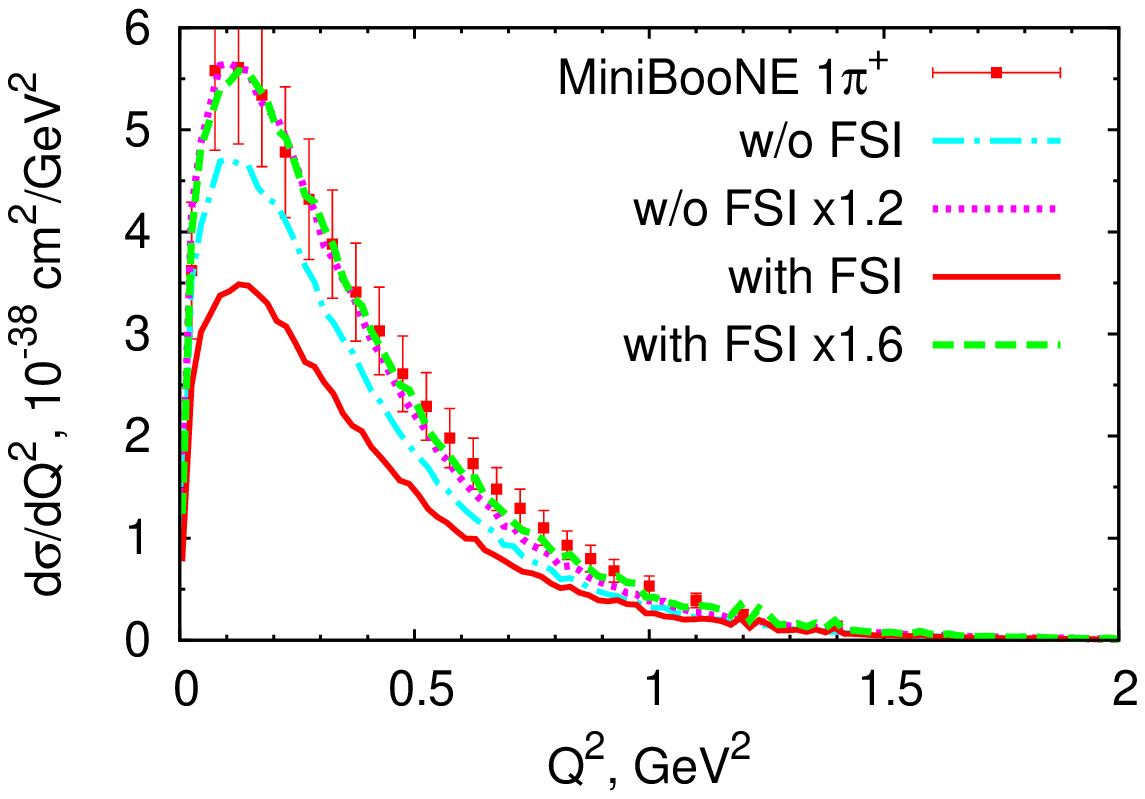}
\end{minipage}
\hfill
\begin{minipage}[c]{0.48\textwidth}
\includegraphics[width=\textwidth]{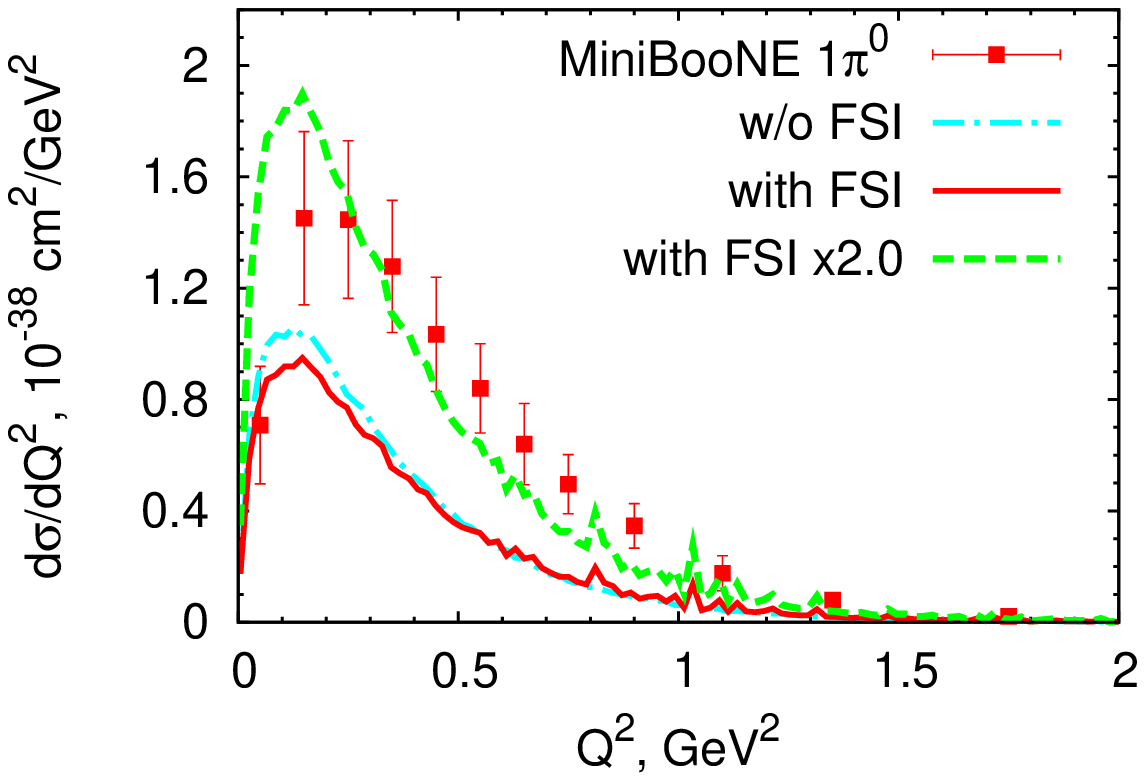}
\end{minipage}
\caption{(Color online) $Q^2$ distribution for MiniBooNE CC 1$\pi^+$(left panel) and 1$\pi^0$
(right panel) production.
Data are from~\protect\cite{AguilarArevalo:2010bm,AguilarArevalo:2010xt}.}
\label{fig:MB-lepton-Q2}

\end{figure}

A more detailed information on pion production processes is contained  in differential distributions.
In Figs.~\ref{fig:MB-lepton-Ekin}, \ref{fig:MB-lepton-Q2} 
we compare the recent MiniBooNE distributions versus muon kinetic energy, $d\sigma/dT_{\mu}$,
and squared four-momentum transfer, $d\sigma/dQ^2$ \cite{AguilarArevalo:2010bm} with our calculations
averaged over the MiniBooNE flux.

As could be expected from the comparison with the  integrated cross sections, our calculations with FSI (solid linse)
and even without FSI (dash-dotted line) are also significantly lower than the experimental data
for both $1\pi^+$ and $1\pi^0$ events.
As in the case of the integrated cross sections, our calculations even without FSI, 
shown as dashed-dotted lines, are about 20\% lower than the data.
These results represent essentially only the free cross section folded with the Fermi momentum-distribution.
The disagreement is therefore, not easy to understand.

To make a shape only comparison for $1\pi^+$, we multiply our ``with FSI'' calculations by a factor 1.6 (long-dashed line)
and the ``without FSI'' by 1.2 (dotted line). These curves practically coincide, which means, that
FSI hardly change the shapes of the distributions. The shape is in good agreement with the shape of the
experimental data for the $Q^2$ distribution, and in reasonable agreement for the kinetic energy distribution.
Recall here that for the integrated cross section we had a noticeable shape disagreement in this channel.

For the $1\pi^0$ channel we we multiply our ``with FSI'' calculations by a factor 2.0 (long-dashed line).
Making a shape-only comparison, one observes noticeable deviations for both
differential cross sections, despite a good shape-only agreement for the integrated cross sections.
Our curve for the $T_\mu$ distribution is noticeably flatter than the data. The calculated $Q^2$ distribution 
is steeper. Since both low  $T_\mu$ and low $Q^2$ originate from low-energy  muons, the disagreement is hard to explain.

\section{Change of the pion spectra due to FSI}

\begin{figure}[!hbt]
\begin{minipage}[c]{0.48\textwidth}
\includegraphics[width=\textwidth]{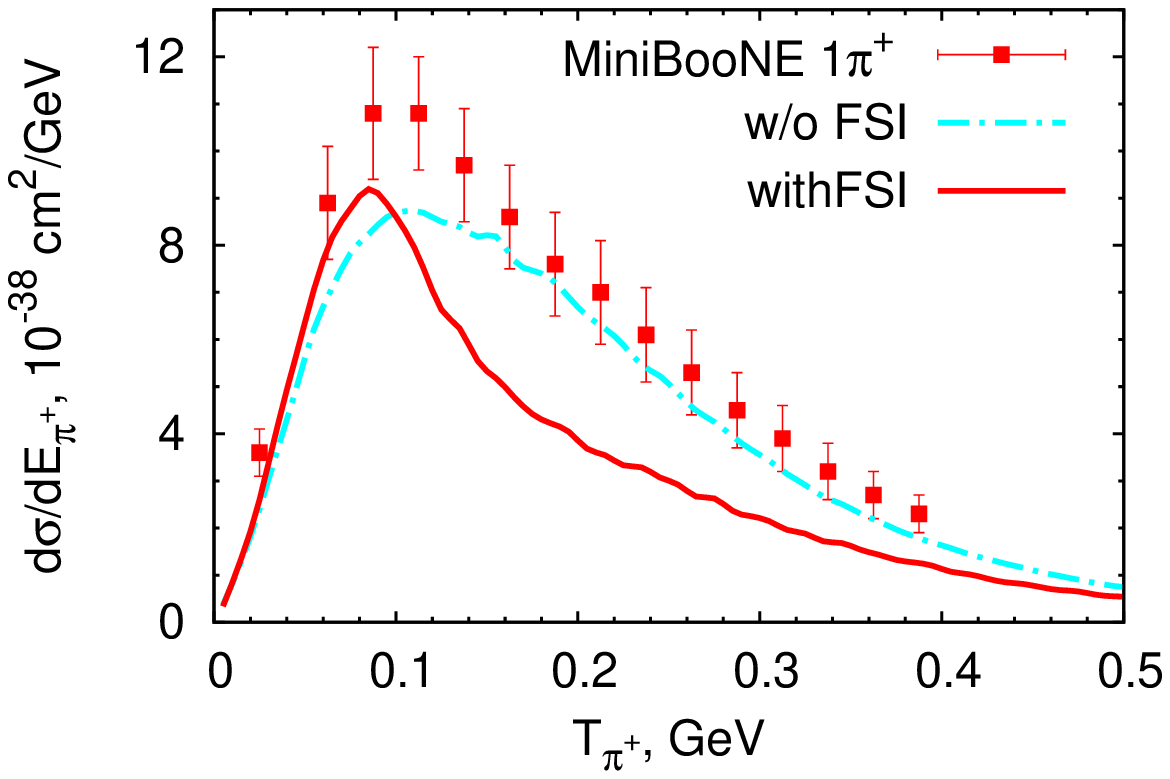}
\end{minipage}
\hfill
\begin{minipage}[c]{0.48\textwidth}
\includegraphics[width=\textwidth]{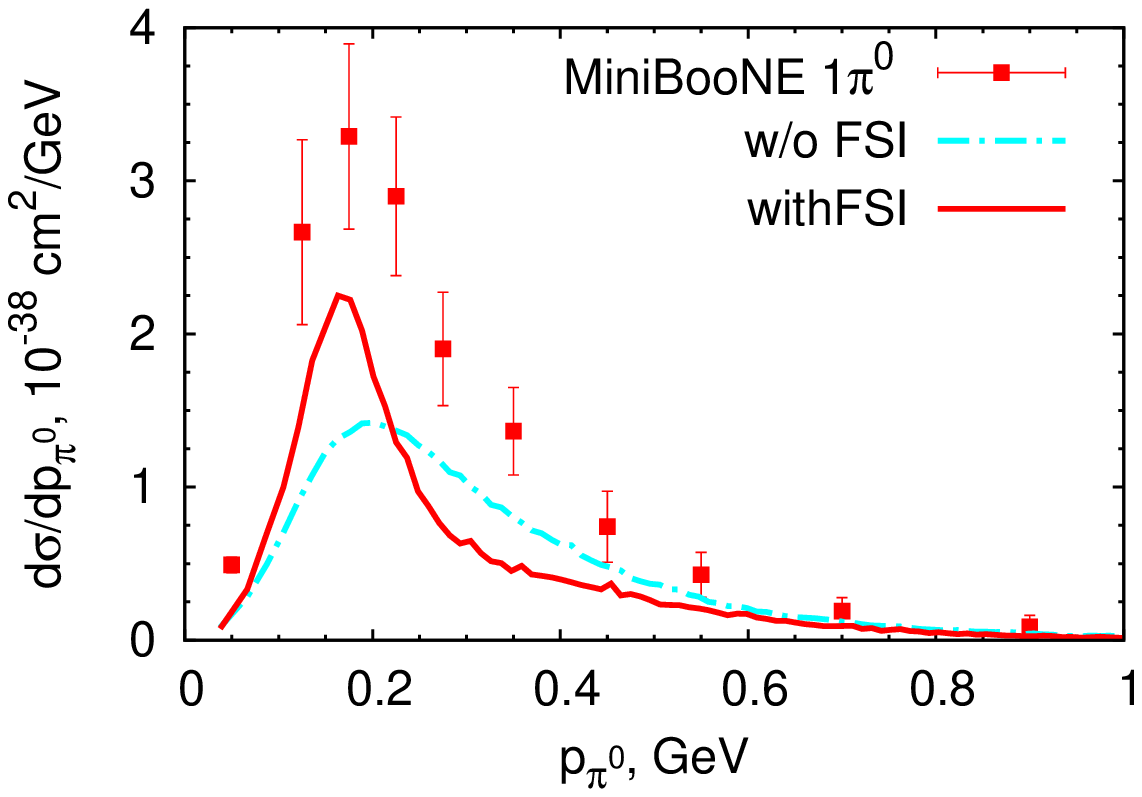}
\end{minipage}
\caption{Distribution of (left panel) the outgoing $\pi^+$ in their kinetic energy; (right panel)
the outgoing $\pi^0$ in their absolute value of the 3-momentum  for the MiniBooNE CC neutrino scattering.
Data are from~\protect\cite{AguilarArevalo:2010bm,AguilarArevalo:2010xt}.}
\label{fig:MB-pion-dTkin}
\end{figure}

Fig.~\ref{fig:MB-pion-dTkin} shows our calculations for the kinetic energy distribution of the outgoing pions. 
As for the other distributions, our calculations with FSI are lower than the experimental data by a factor of 
$1.6$ for $\pi^+$ and a factor of $2$ for $\pi^0$.

A distinctive feature of our calculations is the conclusion, that FSI should significantly
change the pion spectra.
The significant lowering of the cross section for $T_{\pi^+} > 0.12 \GeV$ 
and for $p_{\pi}^0 > 0.25 \GeV$, is a direct consequence of the
pion absorption through $\pi N \to \Delta$ with the following $\Delta N \to N N$.
Pion elastic scattering in the FSI also  decreases the pion energy,
thus depleting spectra at high energies 
and accumulating strength at low energies.
For $\pi^0$ production an additional increase of the cross section at lower energies
comes from the side feeding from the dominant $\pi^+$ channel as discussed above. 
The change of the shape of the spectra is similar to that calculated for
neutral current 1$\pi^0$ production in \cite{Leitner:2008wx}. 

The predicted low-energy peak in pion spectra is 
also similar to that observed experimentally in $(\gamma, \pi^0)$
production on nuclear targets \cite{Krusche:2004uw,Mertens:2008np}. 
Its absence in the neutrino data is, therefore, hard to understand.
At very low energies our curves rise as steeply as the data. After the peak
is reached, however, our calculations predict a rather steep fall--off,
while the data show only a moderate decrease. In this view, it would be 
extremely important to compare our calculations with the coming results from the CLAS experiment
for pion distributions in  electron scattering on nuclei  (see the talk of
S. Manly at these conference).

\section{Any reasonable explanation for discrepancies?}

One possible explanation for the discrepancies observed could be too low elementary cross
sections since the ANL and BNL data for pion production already differ, with
the BNL data being higher by about 30\% (recall that our calculations use
form factors for the $N \Delta$ transition that were fitted to the ANL data).
However, even such an increase would not be sufficient to explain the large
discrepancy by 60\% and more.  

Another source of uncertainties is the
determination of the neutrino energy which is done using quasi-free kinematics
for on-shell Delta production on a nucleon at rest. This, however, have no
influence of flux-averaged observables. One more point is that the experimental data
are presented after being corrected for the finite detection thresholds for muons,
pions and nucleons. This may introduce additional dependence on a particular neutrino 
event generator used in a given experiment \cite{Leitner:2010kp}.

Another possible explanation for the discrepancy is an uncertainty about
the actual reaction mechanism. Recently, several authors have raised the
possibility -- in connection with QE scattering -- that the
impulse approximation,
in which the interactions happen only with one nucleon at a time, may not account for the total cross section and
that instead interactions of the incoming neutrino with nucleon pairs
are important. Indeed, first results
\cite{Martini:2010ex,Martini:2009uj,Amaro:2011qb,Nieves:2011pp}
seem to show that the observed high QE scattering cross sections
could possibly be explained by adding these $2p2h$ contributions
in the primary interaction to the impulse approximation cross section. 
Taking these contributions into account would increase the calculated 
QE-like cross section and thus worsen the agreement for the $\pi$/QE ratio. 
Only when one assumes that a similarly strong contribution  also to pion 
production the ratio would stay unaffected. However, such an increase by about 25\%
 would still be too low to explain
the observed dramatic discrepancy of 60\% between the GiBUU calculations
and the experimental values. 

Another possible explanation for the discrepancy could, of course,
be a too low neutrino flux assumed in obtaining these pion data, 
because that would affect both QE scattering and pion production.
In this view it is interesting to notice, that the agreement between
our calculations and data is much better for flux-independent and
shape-only observables, than for the absolute values.

\begin{theacknowledgments}
This work is supported by DFG. O.L. is grateful to Ivan Lappo-Danilevski for programming assistance.
\end{theacknowledgments}

\bibliographystyle{aipproc}
\bibliography{nuclear}

\end{document}